# Thin film synthesis of SrZn$_2$P$_2$ with SrI$_2$ post-annealing for enhanced crystallinity and optoelectronic quality

Sita Dugu[1], Shaham Quadir[1], Christopher P. Muzzillo[1], Zhenkun Yuan[2,3,4], Smitakshi Goswami [2,5], Xiaojing Hao[6], Jialiang Huang[6], Guillermo Esparza[7], Baptiste Julien[1], David Fenning[7], Jifeng Liu[2], Geoffroy Hautier[2,3,4], Andriy Zakutayev[1], Sage R. Bauers[1]

[1]Materials Sciences Center, National Laboratory of the Rockies, Golden, Colorado 80401, United States
[2]Thayer School of Engineering, Dartmouth College Hanover, NH 03755, USA
[3]Department of Materials Science and Nano Engineering, Rice University, Houston, TX, USA
[4]Rice Advanced Materials Institute, Rice University, Houston, TX, USA
[5]Department of Physics and Astronomy, Dartmouth College, Hanover, NH 03755, USA
[6]Australian Centre for Advanced Photovoltaics, School of Photovoltaic and Renewable Energy Engineering, University of New South Wales, Sydney, New South Wales, Australia
[7]Department of Chemical and Nano Engineering, University of California San Diego, CA 92093, US

## Abstract

Ternary Zintl phosphides are promising light-absorbing semiconductors for thin-film optoelectronic applications, but strategies for controlling their microstructure and optoelectronic quality remain underexplored. Here, we report the synthesis of phase-pure SrZn$_2$P$_2$ thin films using radio-frequency co-sputtering in a PH$_3$ + Ar atmosphere and investigate the impact of post-growth processing on their structural and optical properties. Grazing-incidence X-ray scattering and Raman spectroscopy confirm the formation of crystalline SrZn$_2$P$_2$ films over a finite compositional window. Optical measurements reveal strong absorption near the direct-band-gap energy (~1.8 eV) and near-band-edge photoluminescence. Further, we have studied the effects of chemically compatible halide-assisted annealing. It is found that SrI$_2$ treatments lead to pronounced grain growth and reduced diffraction peak broadening while preserving phase purity, in contrast to rapid thermal or forming-gas annealing. Notably, annealing with SrI$_2$ at 450°C significantly enhances both the intensity and spatial uniformity of the photoluminescence, thus connecting the observed microstructural consolidation with improved radiative recombination. Our study demonstrates that halide-assisted annealing provides an effective pathway for microstructural control in SrZn$_2$P$_2$ thin films and highlights a generalizable processing strategy for advancing Zintl phosphide semiconductors toward optoelectronic applications.

## Introduction

Inorganic thin-film photovoltaic absorbers offer a compelling pathway toward scalable solar energy conversion by enabling strong optical absorption in micron-scale layers and compatibility with diverse deposition platforms. Compared to wafer based technologies such as crystalline Si, thin-film absorber technologies allow for substantial reduction in embodied energy, embodied carbon,[1] material usage and cost while maintaining high performance, as well as new device geometries such as tandem cells.[2–5] These advantages have been demonstrated by mature thin-film technologies such as CdTe and Cu(In,Ga)Se$_2$, which achieve high power conversion efficiencies with absorber layers only 1–2 μm thick.[6,7] However, the reliance of these materials on toxic or supply-constrained elements motivates continued exploration of alternative inorganic thin-film semiconductors.[8–10] Earth-abundant, non-toxic alternatives also based on zinc-blende-derived structures, such as $Cu_2ZnSn(S,Se)_4$ and related compounds, have thus far struggled to achieve high optoelectronic performance.[11,12] On the other hand, while halide perovskites are now a top performing thin film absorber platform, their long-term stability continues to be an issue.[13] Collectively, these limitations highlight the need for new classes of inorganic thin-film absorbers whose chemistry, and microstructural evolution can be systematically understood and controlled.[8]

Recently, high-throughput computational screening identified the ternary Zintl phosphides (e.g., $BaCd_2P_2$) as a class of light-absorbing semiconductors, combining tunable band gaps, strong optical absorption, and attractive defect properties while containing only earth-abundant elements.[14,15] Experimental studies have supported this discovery in $BaCd_2P_2$, which exhibits bright PL, long carrier lifetimes and excellent thermal stability in both bulk and nanocrystalline forms,[14,16,17] as well as in compositionally related $CaZn_2P_2$ thin films, where low-temperature synthesis yields polycrystalline absorbers with carrier lifetimes of tens of nanoseconds.[18] Additionally, $CaCd_2P_2$ has been shown to be a promising visible light-absorbing semiconductor for solar-to-hydrogen energy conversion. Unlike many traditional low-bandgap materials that suffer from photocorrosion, this material exhibits a unique light-induced surface transformation that stabilizes it during the oxygen evolution reaction in alkaline environments.[19]

$SrZn_2P_2$ is a chemically and structurally related member of the Zintl phosphide family, crystallizing in a trigonal structure with a theoretically predicted fundamental indirect band gap of 1.5 eV (1.74 eV direct) making it well suited for visible-light absorption.[15] Despite these attractive intrinsic properties, thin-film synthesis and microstructural optimization of $SrZn_2P_2$ remains entirely unexplored, particularly with respect to developing processing strategies that mitigate grain-boundary-mediated recombination and charge transport. The grain size and crystallinity of the absorber layer play critical roles in charge transport and photovoltaic performance, as larger grains reduce grain boundary recombination and improve carrier mobility, diffusion length, and carrier lifetime.[20] In established thin-film absorbers such as CdTe, post-deposition halide treatments (e.g., $CdCl_2$) have proven essential for promoting grain growth and significantly improving optoelectronic quality, following decades of process refinement.[21–26] Inspired by this precedent, the development of chemically compatible halide-based post-growth treatments represents a promising pathway for functionalizing $SrZn_2P_2$ thin films and improving the broader potential of this emerging absorber family.

In this study, we synthesize $SrZn_2P_2$ thin films, and study post-growth processing in controlling microstructure and optoelectronic properties. Phase-pure $SrZn_2P_2$ thin films are synthesized using $PH_3$-assisted radio-frequency co-sputtering and characterized with X-ray diffraction and optical spectroscopies. We systematically compare several thermal annealing procedures including halide treatments based on chemically compatible $SrI_2$ to evaluate their effects on crystallinity, grain size, and photoluminescence behavior. Our analysis reveals brighter PL emission and up to 200% increase in crystalline grain size after halide treatment. By correlating the processing conditions with structural and optical responses, this work establishes halide-assisted microstructural control as an effective strategy for improving the optoelectronic quality of Zintl phosphide thin films.

## Results

***Thin-Film Synthesis of $SrZn_2P_2$.***

SrZn$_2$P$_2$ thin films were prepared by radio-frequency (RF) co-sputtering onto Si substrates. See the methods section for more details. The resulting thin films exhibit the expected trigonal crystal structure previously observed in bulk material,[27] with Sr ions octahedrally coordinated by P and Zn ions tetrahedrally coordinated by P (Figure 1a). Synchrotron grazing-incidence wide-angle X-ray scattering (GIWAXS) measurements of stoichiometric films grown on Si substrates confirm the formation of polycrystalline SrZn$_2$P$_2$ (Figure 1b). Le Bail refinement against the reported trigonal $P\bar{3}m1$ structure yields lattice parameters of $a$ = 4.117 Å and $c$ = 7.151 Å with a low residual, consistent with the expected crystal structure. The diffraction image of this pattern is shown in figure S1a.

Integrated GIWAXS patterns collected across compositionally graded films (Figure S1b,c) indicate that single-phase SrZn$_2$P$_2$ is maintained over a compositional window of approximately ±3% in either Sr, Zn, or P relative to the ideal stoichiometry. Films with compositions outside this window exhibit either degraded crystallinity or amorphization, illustrating a finite but appreciable tolerance to off-stoichiometric growth conditions. We note that the high crystallinity of SrZn$_2$P$_2$ grown on Si is beneficial for tandem cell applications, as envisioned by Quadir et al.[18] SrZn$_2$P$_2$ films were also deposited on amorphous SiO$_2$ substrates. These samples exhibit a slightly larger compositional spread when compared to the graded films grown on crystalline Si, along with a narrower compositional window of crystallinity (Figure 1c,d). In particular, Sr-rich and Zn-poor compositions tend to form amorphous films, as evidenced by the absence of distinct diffraction peaks.

Raman spectroscopy further confirms the formation of crystalline SrZn$_2$P$_2$ for compositions exhibiting diffraction peaks. As shown in Figure 1e, crystalline films display well-defined E$_g$ and A$_{1g}$ modes, consistent with the calculated vibrational signatures of trigonal SrZn$_2$P$_2$. Note that the two strong and narrow peaks at lower wavenumbers are the primary modes, while those broader ones are 250 and 300 cm$^{-1}$ are higher order modes. As shown in Table S1, the experimental Raman frequencies are close to our first principles calculations. In contrast, Raman features are absent in amorphous compositions, corroborating the loss of long-range order inferred from X-ray diffraction.

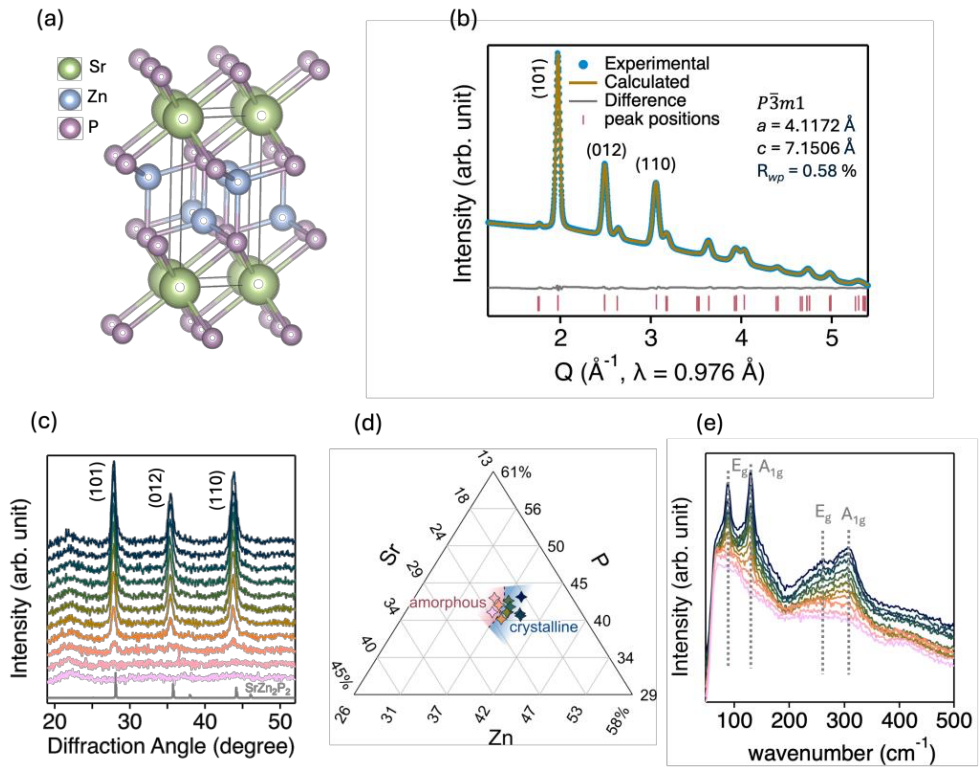

Fig 1: (a) Crystal structure of trigonal SrZn$_2$P$_2$ (b) GIWAXS X-ray diffraction (XRD) of stoichiometric SrZn$_2$P$_2$ film with Le Bail fit against the trigonal $P\bar{3}m1$ space group. (c) XRD patterns of 11 compositionally graded points across a film, with reference pattern (ICSD #30911) at the bottom. (d) X-ray fluorescence reading of the corresponding XRD patterns. (e) Raman spectra of corresponding points. Colors are consistent across panes c–e.

To evaluate the presence of oxygen and other elemental impurities, X-ray photoelectron spectroscopy (XPS) was also performed on neat surface and on Ar-etched surface (see Figures S2). The XPS spectra show evidence of the formation of carbonates on the air-exposed surface, as reported for CaZn$_2$P$_2$.[18] However, the carbonate C 1s and O 1s signals significantly diminish after etching with Ar gas clusters, indicating that oxidation/carbonate formation is mostly limited to the surface of the films.

*Optical Properties of Thin-Film SrZn$_2$P$_2$*
The optical properties of SrZn$_2$P$_2$ thin films were evaluated using spectroscopic ellipsometry. Figure 2a shows the refractive index ($n$), extinction coefficient ($k$), and absorption coefficient ($\alpha$) extracted from ellipsometry measurements. SrZn$_2$P$_2$ exhibits an absorption onset near 1.5 eV, consistent with its fundamental band gap.[15] The pre-edge region is shaded in blue. A change in curvature of the absorptivity curve between 1.5–2 eV (red shaded region) may correspond to additional absorption across the direct gap and several other conduction band minima (discussed later). Overall, a strong absorptivity of $1.2 \times 10^4$ cm$^{-1}$ is observed by 1.8 eV. Details of the ellipsometry modeling (Figure S3a) and band gap analysis are provided in the Supporting Information. Analysis of the absorption edge using the Tauc plot approach indicates an indirect fundamental gap near 1.5 eV (Figure S3b), consistent with electronic structure calculations for SrZn$_2$P$_2$[15] (also discussed below in figure 5). The magnitude and sharpness of the absorption edge indicate that $SrZn_2P_2$ is a strong visible-light absorber suitable for thin-film optoelectronic applications.

The optical absorptivity of $SrZn_2P_2$ thin films near stoichiometry (with the color corresponding to the compositions shown in Figure 1d) was measured by UV-vis spectroscopy at room temperature, shown in Figure S4a. Similar to ellipsometry measurement, strong absorption with $\alpha > 10^4$ cm$^{-1}$ within 100 meV above band edge is observed. No substantial variation in absorption strength is observed between films at crystalline and amorphous compositions, indicating that optical absorption in $SrZn_2P_2$ is relatively insensitive to long-range crystallinity and few-percent deviations from stoichiometry. The reflectance, transmittance and absorbance of white light is shown in figure S4b. Interference fringes observed in reflectance and transmittance spectra indicate smooth film surfaces/interfaces and a uniform thickness, and explain the oscillations observed in the absorptivity curves.

Cathodoluminescence (CL) spectroscopy provides confirmation of radiative recombination in $SrZn_2P_2$. Figure 2b shows a plan-view scanning electron micrograph (SEM) of a $SrZn_2P_2$ thin film, as well as uniform CL signal collected from the film. Normalized CL intensity vs. energy spectra show the emission peak is centered at 1.78 eV, shown in figure 2c, in close agreement with the calculated 1.74 eV direct band gap[15] and photoluminescence (PL) measurements discussed below. Energy-dispersive X-ray spectroscopy (EDS) elemental mapping performed at the same locations as the CL measurements confirms compositional uniformity (Figure S5), supporting the intrinsic origin of the observed luminescence.

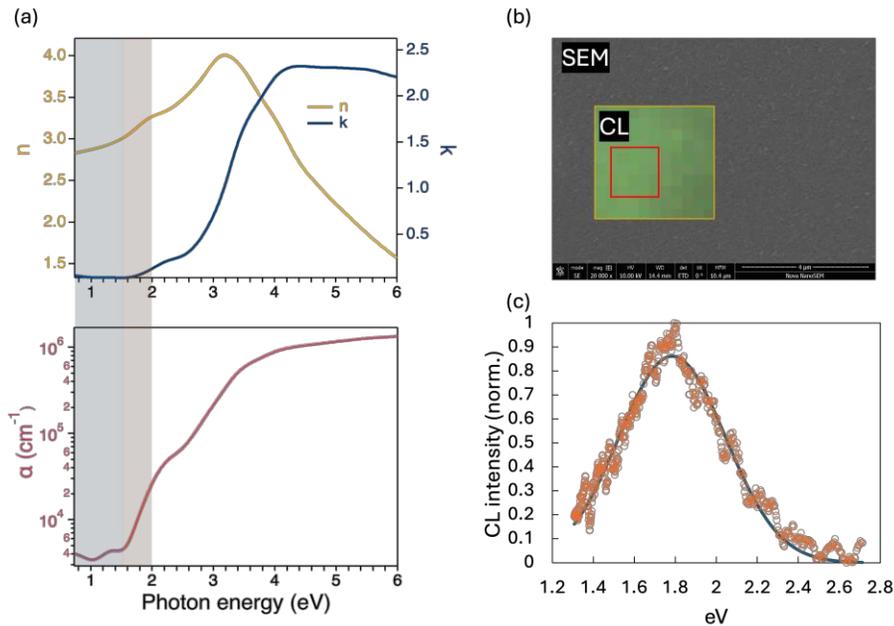

Figure 2: (a) Refractive index (n), extinction coefficient (k), and absorption coefficient ($\alpha$) plot versus photon energy. (b) Plane-view SEM image and cathodoluminescence image collected from $SrZn_2P_2$. (c) CL spectrum of $SrZn_2P_2$ scanned at red square area in (b).

*Halide-Assisted Annealing*

Optoelectronic performance in polycrystalline semiconductors like $SrZn_2P_2$ is often limited by extrinsic factors such as extended defects and microstructural disorder. Grain enlargement through recrystallization in fluxes is a well-established strategy to improve carrier lifetime and charge transport in polycrystalline thin-film solar absorbers, namely CdTe,[7,21,28] as grain boundaries introduce recombination-active defect states.[29] This motivates the study of annealing treatments to enhance the microstructural quality of $SrZn_2P_2$ thin films.

To benchmark conventional thermal processing without the aid of a flux, $SrZn_2P_2$ films were annealed by rapid thermal annealing (RTA) at 350°C and 500°C. Plan-view SEM images (Figures S6a-c) reveal minimal changes in grain size (~100-150 nm), although grain boundaries become more clearly defined after annealing. Figure S6d is diffraction patterns of as-deposited and annealed $SrZn_2P_2$ films, which show a slight reduction in diffraction peak full width at half maxima (FWHM) with increasing temperature (Figure S6e), consistent with modest improvements in crystalline quality. However, the FWHM of the 350°C-annealed film is within the pristine film's error window and annealing at 500°C leads to the emergence of a carbonate-related diffraction feature near $2\theta \approx 34°$ (Figure S6d). Given carbonate was only observed by XPS on the surface of the films before annealing (*cf.* Figure S2), this emergent diffraction signal suggests deeper penetration into the film and side reactions from imperfect purge conditions. These results indicate that conventional thermal annealing alone is insufficient to drive significant grain growth in the $SrZn_2P_2$ thin films before side reactions occur.

Halide salts are often employed as recrystallization fluxes in thin-film absorbers. The most notable example is $CdCl_2$ processing of CdTe.[21,26] In $SrZn_2P_2$, however, direct application of an analogous $ZnCl_2$ salt is likely to lead to the double ion-exchange reaction $SrZn_2P_2 + ZnCl_2 \rightarrow Zn_3P_2 + SrCl_2$, driven by the large heat of formation for $SrCl_2$ ($\Delta H$ (s) = –829 kJ/mol) over $ZnCl_2$ ($\Delta H$ (s) = –415 kJ/mol).[30] The formation of $PX_3$ ($X$ = halide) compounds, which would lead to phosphorous loss, must also be considered. The enthalpies of formation for $PX_3$ compounds are, $\Delta H$ = –320 kJ/mol for $PCl_3$ (l), $\Delta H$ = –185 kJ/mol for $PBr_3$ (l), and $\Delta H$ = –46 kJ/mol for $PI_3$ (s)[30] indicating the best choice is an iodide. Similar to the chlorides, heats of formation for $SrI_2$ ($\Delta H$ (s) = –558 kJ/mol) and $ZnI_2$ ($\Delta H$ (s) = –208 kJ/mol) make $ZnI_2$ unattractive. Thus, in consideration of available halide salts and the various competing reactions that destabilize the target phase and/or promote phosphorous loss, $SrI_2$ was selected as the most-probable chemically compatible halide salt for post-growth annealing of $SrZn_2P_2$ thin films.

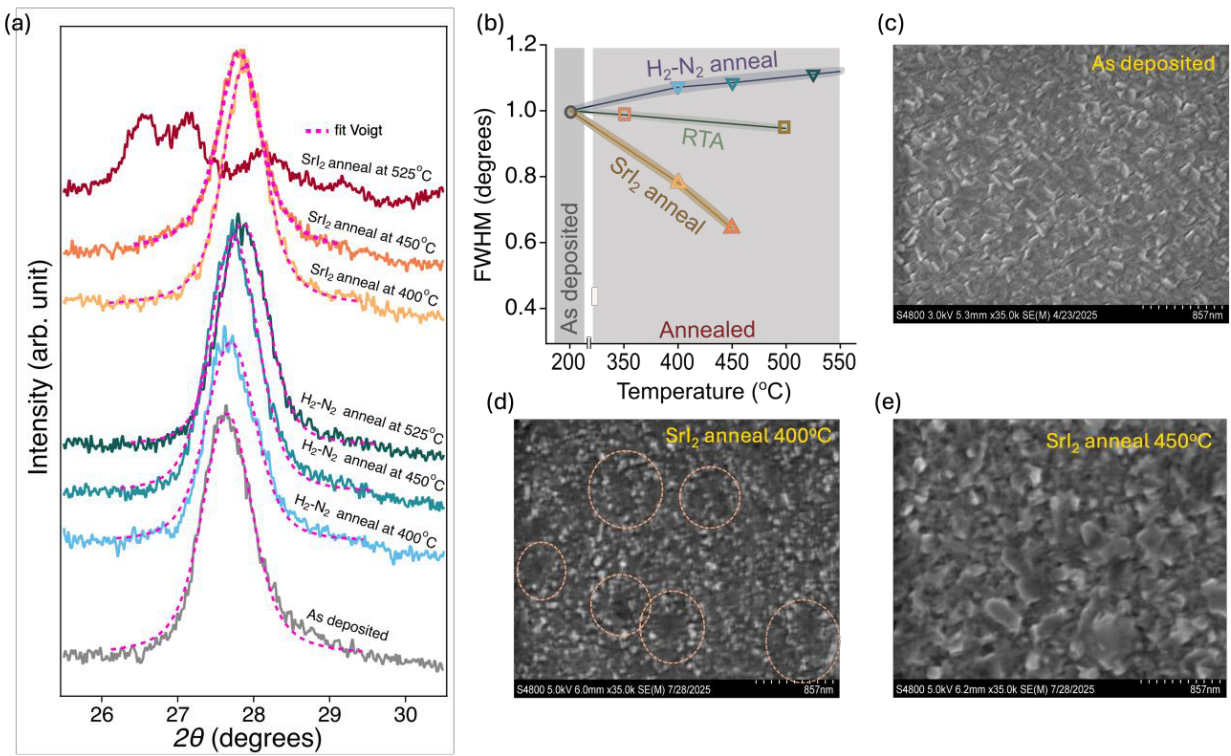

Figure 3: (a) X-ray diffraction data for both as-deposited and annealed $SrZn_2P_2$ thin films, with pseudo-Voigt fits of the (101) reflection. (b) Full-width-at-half-maxima extracted from the fits shown in pane (a), as well as rapid thermal annealed films (see Figure S6 for XRD). (c-e) Plan-view scanning electron micrographs of $SrZn_2P_2$ films as-deposited, $SrI_2$ annealed at 400°C, and $SrI_2$ annealed at 450°C respectively.

$SrI_2$-assisted annealing yields significant microstructural improvements compared to conventional thermal treatments. Figure 3a shows XRD patterns of films both as-deposited and annealed under various conditions. Phase-pure $SrZn_2P_2$ is preserved following $SrI_2$ annealing up to 450°C, while higher-temperature treatment at 525°C with $SrI_2$ leads to degradation. Quantitative analysis of the (101) diffraction peak (Figure 3b) reveals a substantial decrease in FWHM (normalized to the FWHM of the as-deposited film) with increasing $SrI_2$ annealing temperature, indicating improved crystalline coherence. The FHWM is reduced to ~0.6° 2θ after $SrI_2$-450°C anneal. Samples were also annealed in $N_2$ with an $H_2$ pre-anneal ($H_2$-$N_2$ anneal) at identical temperature for comparison. The $H_2$-$N_2$ anneal without the halide salt flux produces no grain enlargement (Figure S7) and instead results in a slight increase in FWHM with each increase in temperature.

Complementary plan-view SEM images in Figure 3c-e show that $SrI_2$ annealing induces pronounced grain growth. While films annealed at 400°C just begin to exhibit agglomeration and coalescence, as highlighted by the dotted circles, the surfaces of samples annealed at 450°C comprise enlarged domains relative to the as-deposited film. While pre-annealed grain features are on the order of 100-150 nm in plan-view SEM, this size increases to 200-300 nm after annealing. These observations indicate that $SrI_2$ acts as an effective fluxing agent, enabling grain-boundary migration and microstructural consolidation at temperatures where conventional annealing is ineffective. Enhancement of SrZn$_2$P$_2$ crystallinity with halide annealing is further confirmed by Raman spectroscopy. Figure S8 shows the Raman spectra for both as-deposited and SrI$_2$-450°C annealed samples. The FWHM of the primary E$_g$ peak decreases from 3.1 cm$^{-1}$ to 1.4 cm$^{-1}$ with annealing. Furthermore, no secondary phases are seen after annealing.

The grain ripening while preserving phase purity in the SrI$_2$-annealed film at 450°C makes this condition suitable for studying how microstructure improves the optoelectronic response. Photoluminescence spectroscopy was used to study optical properties of pre- and post-annealed films at 100 locations, as shown in Figure 4. Data for the as-deposited films are shown in the upper pane (in blue curves) with light blue curves representing the individual 100 PL spectra and the dark blue curve being their average. Luminescence from SrZn$_2$P$_2$ is centered at 1.8 eV corresponding to near-band-edge radiative recombination and in good agreement with the CL measurements shown in Figure 2c. A higher-energy and higher intensity feature at ~2.06 eV is attributed to emission from the amorphous SiO$_2$ substrate.

PL spectra shown in the lower pane (in light green curves) are collected from 100 locations across the same region after SrI$_2$-450°C annealing. It shows highly reproducible emission profiles despite the spatial variation, in contrast to the broader distribution observed for the as-deposited film. The average PL spectrum (dark green curve) reveals enhancement of the near-band-edge emission intensity following SrI$_2$ treatment. There is a slight red shift of the annealed film's emission to 1.74 eV, identical to the calculated direct band gap energy (Figure 5). The enhanced emission is especially clear when comparing the relative emission intensity between SrZn$_2$P$_2$ and the substrate's 2.06 eV color center, which is not expected to change from the annealing process. Normalizing to the 2.06 eV color center intensity, we estimate that the PL intensity of SZP increased by nearly 10-fold after the treatment. Furthermore, we can also observe a transition peak at ~1.4 eV after treatment. An explanation is that the non-radiative recombination is notably reduced as the indirect gap radiative recombination lifetime is typically long and can be easily smeared out by non-radiative defects. This PL enhancement is consistent with improved crystallinity and reduced non-radiative recombination associated with grain growth and microstructural consolidation. Halide-assisted treatments are known to modify grain boundaries in related thin-film absorber systems. Halide ions diffuse into and passivate these boundaries leading to reduced recombination activity,[31–33] and a similar mechanism may contribute here. SrI$_2$'s band gap is wider than the absorber[34] so it may passivate SrZn$_2$P$_2$ surfaces.[35] While the precise microscopic role of the halide salt requires further investigation, the strong correlation between grain enlargement, reduced diffraction peak broadening, and enhanced PL intensity supports the conclusion that halide-assisted microstructural control directly improves the optoelectronic quality of SrZn$_2$P$_2$ thin films.

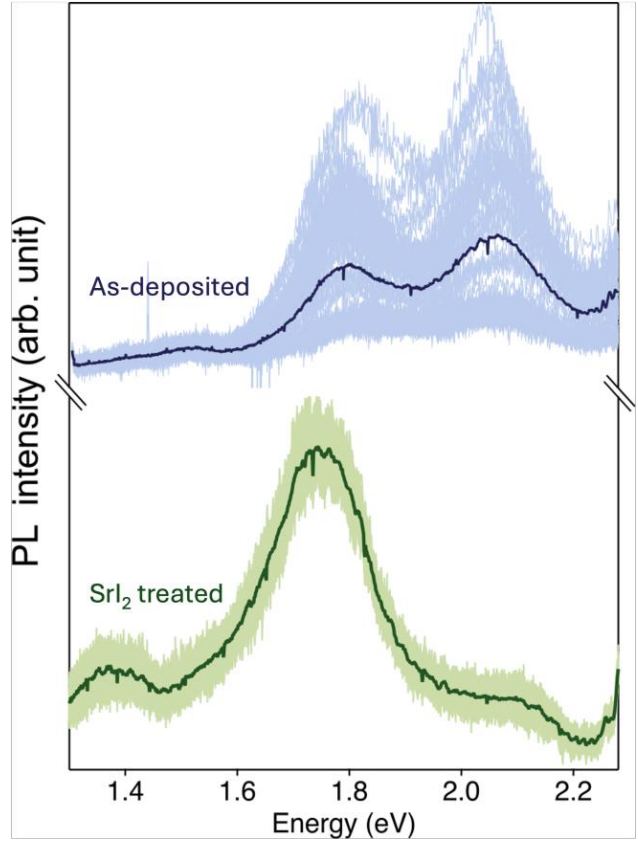

Figure 4: Photoluminescence of the film at scanned at 100 points- upper blue curves are PL for as-deposited film scanned at 100 points, dark blue is average of them. Lower light green curves are PL for SrI$_2$-450°C annealed film scanned at 100 points, dark green is average of them.

*Calculations on SrZn$_2$P$_2$.*

The successful thin film synthesis and improvement of the synthesized SrZn$_2$P$_2$ films through halide salt annealing positions this material as a promising semiconductor provided other properties such as carrier mobility, intrinsic defects and dopability are also favorable. To provide context for the experimentally observed optical behavior of SrZn$_2$P$_2$ and answer these outstanding questions, we have also studied the electronic structure and intrinsic defect properties of SrZn$_2$P$_2$ using hybrid functional calculations. Figure 3a shows the calculated electronic band structure and density of states (DOS), which indicate that SrZn$_2$P$_2$ is an indirect gap semiconductor, with the valence-band maximum (VBM) and conduction-band minimum (CBM) located at the $\Gamma$ and M points, respectively, leading to a 1.50 eV fundamental indirect band gap. The direct band gap, found at the $\Gamma$ point, is 1.74 eV. Despite the indirect nature of the band gap, the small energy difference between the indirect and direct gaps, as well as a high conduction band-edge density of states due to the nearly flat conduction band between the $\Gamma$ and A points, support strong optical absorption near the direct transition, consistent with the experimentally measured absorption and PL/CL. Based on the calculated band structure, we have also evaluated the carrier effective masses using the Boltzmann transport equation.[36] At a doping concentration of $10^{16}$ cm$^{-3}$ and 300 K, we find relatively small conductivity effective masses for both holes (0.35–0.64 $m_0$) and electrons (0.24–1.41 $m_0$).

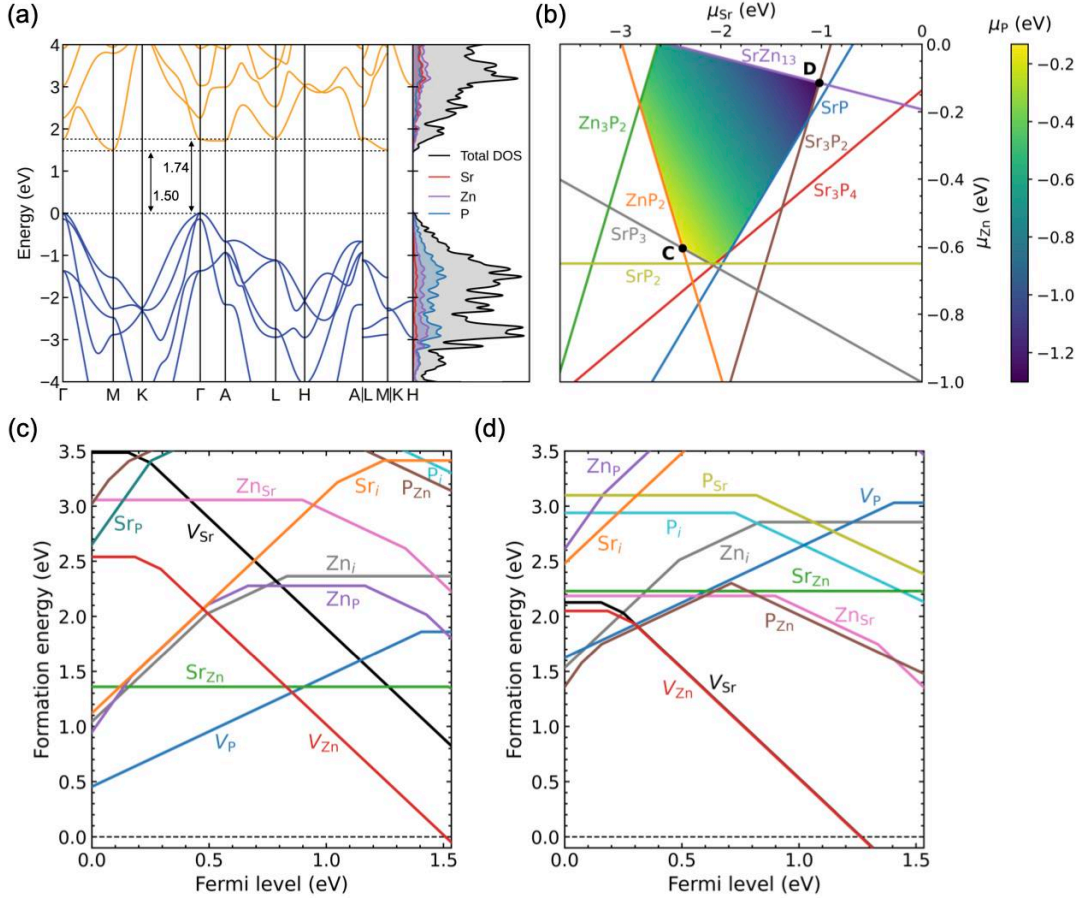

Figure 5: (a) HSE06-calculated electronic band structure and density of states (DOS) of SrZn$_2$P$_2$. (b) HSE06-calculated chemical-potential stable region of SrZn2P2 (colored area). Two chemical-potential points, C and D, are labeled and used to represent the P-poor and P-rich chemical-potential conditions, respectively. (c and d) HSE06-calculated formation energies for intrinsic point defects in SrZn2P2 as a function of Fermi level at the chemical-potential points C and D.

The chemical-potential stable region of SrZn$_2$P$_2$ is shown in Figure 5b. SrZn$_2$P$_2$ exists over a broad phase-width, supporting the experimental observation that several growth conditions and even a few percent of off-stoichiometry can be tolerated before secondary phases or amorphization are observed. A narrow phase width has led to practical problems in other PV materials, most notably Cu$_2$ZnSnS$_4$.[37] The broad stability window for SrZn$_2$P$_2$ spanning >0.6 eV across $\mu_{Sr}$, $\mu_{Zn}$, and $\mu_P$ might thus be a hidden advantage of this material.

Figures 5c and 5d show the calculated formation energies for the intrinsic point defects in SrZn$_2$P$_2$ under P-poor and P-rich conditions (corresponding to chemical-potential points **C** and **D** in Figure 5b), respectively. We find that SrZn$_2$P$_2$ shows benign defect properties, similar to what we previously reported for BaCd$_2$P$_2$ and CaZn$_2$P$_2$.[14,18] Across the representative chemical-potential conditions, only vacancy-type defects, including $V_{Sr}$, $V_{Zn}$, and $V_P$, are predicted with low formation energies, while antisite- and interstitial-type defects are energetically unfavorable, irrespective of growth conditions and the Fermi-level position. Importantly, both $V_{Sr}$ and $V_{Zn}$ are shallow acceptors, while $V_P$ is a shallow donor. Therefore, the defect calculations suggest that deep intrinsic defects are unlikely to form in high concentrations in SrZn$_2$P$_2$, which will be beneficial to the carrier lifetime.

The calculated defect formation energies indicate that SrZn$_2$P$_2$ will be an intrinsic material in the absence of any impurities. Under P-poor conditions (Figure 5c), the $V_{Zn}$ acceptor and the $V_P$ donor will pin the Fermi level close to the intersection of their formation-energy lines (i.e., at $E_F = 0.85$ eV). Therefore, in those conditions, the Fermi level will be far from the band edges. Under P-rich conditions (Figure 5d), the formation energy of $V_P$ becomes much higher. The formation energies of intrinsic $V_{Sr}$ and $V_{Zn}$ acceptors are still high, so it is unlikely that cation vacancies alone will result in a high hole concentration at room temperature. However, due to the high formation energy of $V_P$ and other donor defects, their compensation effect will be weak and SrZn$_2$P$_2$ is expected to be highly p-type dopable. Therefore, if SrZn$_2$P$_2$ is to be used as a p-type semiconductor, it should be prepared using P-rich growth conditions with incorporation of shallow extrinsic p-type dopants.

## Conclusion:

In summary, SrZn$_2$P$_2$ thin films were successfully synthesized by radio-frequency co-sputtering, with phase-pure crystalline material confirmed by GIWAXS and Raman spectroscopy. Optical characterization reveals strong absorption near the direct band gap energy in both crystalline and amorphous films, establishing SrZn$_2$P$_2$ as an optically active thin-film semiconductor. Post-growth annealing in the presence of SrI$_2$ leads to pronounced microstructural evolution that is not achieved through conventional thermal or forming-gas annealing. In particular, SrI$_2$ annealing at 450°C produces significant grain enlargement, reduced diffraction peak broadening, and a substantial enhancement in near-band-edge photoluminescence intensity. These correlated structural and optical improvements indicate that processing with halide salt fluxing agents effectively improves the optoelectronic quality of SrZn$_2$P$_2$ thin films by mitigating microstructure-related non-radiative losses. More broadly, this work demonstrates that chemically informed halide-assisted annealing provides a viable strategy for microstructural control in the broader family of Zintl phosphide thin films, highlighting a promising pathway for advancing this emerging class of inorganic optoelectronic materials toward device applications.

## Methodology

*Thin film synthesis and processing:*
Combinatorial composition gradient SrZn$_2$P$_2$ films were grown by keeping the substrate stationary against the confocally oriented sputter cathodes and the uniform films were obtained by rotating the platen at 20 rpm during the growth. 50.8 mm diameters metallic Sr and Zn targets of power densities 1.13 – 1.25 W cm$^{-2}$ and 2.96 W cm$^{-2}$ respectively were used for RF co-sputter. Depositions were carried out for 2 hours at 200°C in a reactive gas environment composed of 2% PH$_3$ and 98% Ar introduced at the flowing of 19.5 sccm. **WARNING:** Due to the toxicity and pyrophoric nature of PH$_3$, stringent safetly protocols were strictly followed throughout the film growth process. Details regarding the deposition sustem, safely handling PH$_3$ as a process gas and safety precaution protocols during chamber opeartion is described in previous works[18,38]. The process pressure was regulated using a throttled gate valve, with a final working pressure of 5 mTorr corresponding to 0.1 mTorr of PH$_3$. The base pressure was manitained below 1 x 10$^{-7}$ Torr. Once the chamber is cooled to room, it was purged with O$_2$ temperature prior to transfer sample to the loadlock.

For halide-assisted annealing experiments, fresh SrI$_2$ powder was loaded into a graphite boat. The SrZn$_2$P$_2$ thin-film samples were placed 3 mm above the SrI$_2$. The graphite boat and lid provided independent temperature control of the source SrI$_2$ powder and SrZn$_2$P$_2$ film, respectively, in a cold-wall quartz reactor. The source and substrate were prebaked at 200 °C in H$_2$ for 10 min to desorb carbonates, then the system was pumped out and filled with 400 Torr N$_2$. Finally, the substrate and source were ramped to 450 °C and 500 °C, respectively, and held for 30 min. Anneal temperature and time was optimized after after multiple test in different temperature from 400°C to 550°C. Here we chose SrI$_2$ annealing at 400°C for 90 minutes, 450°C for 30 minutes, and 525 °C for 10 minutes.

*Structural characterization:*

Experimental data in this study were analyzed using COMBIgor package,[39] and are publicly available publicly in the National Laboraroty of the Rockies (NLR) high-throughput experimental material database.[40,41] Synchrotron grazing incidence wide angle X-ray scattering (GIWAXS) measurements were performed at beamline 11–3 at the Stanford Synchrotron Radiation Lightsource, SLAC National Accelerator Laboratory. The data were collected with a Rayonix 225 area detector using a wavelength of $\lambda$ = 0.97625 Å, a 1° incident angle, a 150 mm sample-to-detector distance, and a beam size of 50 μm vertical x 150 μm horizonal. The diffraction images were calibrated with a LaB6 standard and integrated with the Nika SAS package. Integrated data were averaged from 5 frames of 15 s each. X-ray diffraction (XRD) data were collected on a Bruker D8 diffractometer using Cu K$\alpha$ radiation and a 2D detector. An X-Y stage was used to provide spatial mapping capability over compositionally graded thin films. Patterns were integrated to generate an intensity vs. 2$\theta$ pattern. Elemental composition Sr, Zn and P were checked by X-ray fluorescence (XRF) using a Rh anode at 50 keV and spectra were modeled as a SrZn$_2$P$_2$ layer with unknown composition and thickness on top of a Si or aSiO$_2$ substrate. Raman measurements were performed on a Renishaw inVia (Gloucestershire, UK) PL/Raman microscope equipped with a 532 nm laser and 20× magnification objective lenses. Grating (1800 lines mm$^{-1}$) was used to direct scattering light from the sample to the CCD detector.

X-ray photoelectron spectroscopy (XPS) were performed using an AXIS- Supra by Kratos Analytical with Au and Cu calibration, using a Al K-alpha photon source. Etching was done using an Ar gas cluster ion source. The films measured were both aged in air for multiple weeks as well as stored in a nitrogen environment immediately after deposition, though these two conditions did not vary significantly. Survey and high-resolution scans of Sr 2p, Zn 2p, Zn 1s-P 2p, C 1s, and O 1s were performed on the films as-is (without cleaning) as well as after a sputtering. A 60 s etch time was deemed appropriate by using 10 s etch cycles and monitoring the strength of the C 1s signal. XPS peak fitting was done in CasaXPS using Tougaard and Shirley backgrounds. Lorenztian asymmetric lineshapes were used to fit the peaks, with scans to optimize the lineshape parameters.

*Optical charaterization:*

Spectroscopic ellipsometry data were acquired using a J. A. Woollam Co. M-2000 variable angle ellipsometer at incident angles of 65°, 70°, and 75° over a photon energy range of 0.73–6.46 eV. Raw $\Psi$ and $\Delta$ data were modeled and fit with B-Spline model using the CompleteEASE software (version 6.63) to extract optical parameters n, k, and absorption coefficient. Optical transmission (T) and reflection (R) spectra were also collected in the ultraviolet and visible (UV–vis) spectral regions on a custom-built optical spectroscopy system. A blank substrate was measured as a perfect transmission standard immediately before the SrZn$_2$P$_2$ film. An Al mirror was similarly used as a perfect reflection standard. The optical absorption (A) was determined from the fact that A+T+R = 100%. The absorption coefficient, $\alpha$, was determined by numerically fitting the R, T data using transfer matrix method, as discussed in the main text.

Micro-photoluminescence (PL) and Raman spectra of $SrZn_2P_2$ thin films deposited on amorphous $SiO_2$ substrates were acquired at room temperature using a confocal micro-PL system (Horiba LabRAM HR Evolution). A 532 nm continuous wave diode pumped solid-state laser (nominal output 100 mW) served as the excitation source. The laser beam was directed through a 50× long-working-distance objective (NA = 0.50), producing a spot size of ~3 μm diameter on the film surface. The emitted PL was collected by the same objective and dispersed by a 300 grooves/mm grating. To distinguish film emission from substrate-induced signals, PL measurements were repeated on a bare $SiO_2$ substrate under identical excitation and collection conditions. From these data, the characteristic $SiO_2$ color-center emission at ~2.1 eV was identified and subsequently included when interpreting the $SrZn_2P_2$ thin-film spectra. A rectangular region (x = −63.53 to 65.90 μm, y = −57.87 to 52.97 μm) was mapped using a uniform spatial grid with step sizes of 14.6 μm (x) and 13 μm (y) and a probe spot size of 10 μm, schematic shown in figure S9. In total, 100 spatially resolved measurements were acquired for the as deposited and the $SrI_2$ treated thin film samples, with a total acquisition time of approximately 4 h 15 min. The schematic of 100 points scans, and their point distance is shown in figure S9. Neutral density (ND) filters of 3.2% and 5% were used for PL and Raman measurements, respectively. The higher excitation power used for Raman compensates for the intrinsically weak scattering signal while avoiding laser-induced heating and spectral distortion. Raman spectra were acquired at discrete points rather than via spatial mapping.

Plan-view CL analysis was conducted using Delmic SPARC spectral cathodoluminescence system coupled to an FEI Nova Nano SEM 450 field-emission SEM. Measurements were performed at an accelerating voltage of 15 kV with a beam current of ~0.8 nA, and the dwell time was set to 5s per pixel. Scanning electron micrograph (SEM) images were analysed on the films deposited on Si substrate using a Zeiss Sigma 500 microscope, operating at a 3.00 kV accelerating voltage, using an in-lens secondary electron detector. No conductive coating or additional modification was performed on the samples.

*Theoretical calculations*

The first-principles calculations were performed using the projector augmented-wave (PAW) method and the HSE06 hybrid functional as implemented in the Vienna ab initio simulation package (VASP).[42–45]. The pseudopotentials for Ca (Ca_sv), Zn, and P from the PBE PAW datasets (version 54) were used. An energy cutoff of 400 eV was used for the plane-wave basis set. Using an 8 × 8 × 4 $\Gamma$-centered **k**-point grid, the lattice constants of $SrZn_2P_2$ are found to be: $a = b = 4.099$ and $c = 7.102$. The calculations of intrinsic point defects in $SrZn_2P_2$ were performed using a 240-atom supercell, which is a 4 × 4 × 3 repetition of the relaxed unit cell. For supercells containing a defect, all the internal atomic positions were fully relaxed until the interatomic forces become smaller than 0.01 eV/Å. For the defect calculations, a $\Gamma$-only **k**-point grid was used. Spin polarization was explicitly included in all the defect calculations. The defect formation energies were computed using the standard formalism,[46] with total energy corrections for charged supercells.[47] The chemical-potential stable region of $SrZn_2P_2$ was determined using the calculated formation enthalpies, which are −2.27, −2.55, −2.78, −5.67, −7.16, −1.22, −0.87, and −2.51 eV/f.u. for $SrP$, $SrP_2$, $SrP_3$, $Sr_3P_2$, $Sr_3P_4$, $Zn_3P_2$, $ZnP_2$, and $SrZn_{13}$, respectively. The open-source software PyCDT [48] and Pydefect[49] were used in the defect calculations. The Python toolkit sumo[50] was used for visualizing the electronic band structure and DOS.

## Supporting Information

The data supporting this article have been included as part of the ESI.

## Acknowledgements

This work was authored in part by the National Laboratory of the Rockies, operated by Alliance for Advanced Energy, LLC, for the U.S. Department of Energy (DOE) under contract no. DE-AC36-08GO28308. Funding provided by the U.S. Department of Energy, Office of Science, Basic Energy

Sciences, Division of Materials Science and Engineering, Physical Behavior of Materials program through the Office of Science Funding Opportunity Announcement (FOA) Number DE-FOA-0002676: Chemical and Materials Sciences to Advance Clean-Energy Technologies and Transform Manufacturing under award number DE-SC0023509. This research used resources of the National Energy Research Scientific Computing Center (NERSC), a DOE Office of Science User Facility supported by the Office of Science of the U.S. Department of Energy under contract no. DE-AC02-05CH11231 using NERSC award BES-ERCAP0023830. Use of the Stanford Synchrotron Radiation Lightsource, SLAC National Accelerator Laboratory, is supported by the U.S. Department of Energy, Office of Science, Office of Basic Energy Sciences under Contract No. DE-AC02-76SF00515. The authors acknowledge the use of facilities and instrumentation at the UC Irvine Materials Research Institute (IMRI), which is supported in part by the National Science Foundation through the UC Irvine Materials Research Science and Engineering Center (DMR- 2011967). *Thanks to Dr. Ich Tran for support with photoelectron spectroscopy data collection.* This work was performed in part at the San Diego Nanotechnology Infrastructure (SDNI) of UC San Diego, a member of the National Nanotechnology Coordinated Infrastructure, which is supported by the National Science Foundation (grant ECCS-2025752). The views expressed in the article do not necessarily represent the views of the DOE or the U.S. Government.

## Author Contribution

Sita Dugu: conceptualization, investigation, methodology, validation, writing - orginal draft. Shaham Quadir: investigation, writing - review & editing. Christopher P. Muzzillo: investigation, methodology, writing - review & editing. Zhenkun Yuan: methodology, software, visualization, Writing - original computational draft. Smitakshi Goswami: investigation, writing -review &editing. Xiaojing Hao: investigation, visualization, writing - review & editing. Jialiang Huang: investigation, visualization, writing - review & editing. Guillermo Esparza : investigation, visualization, writing - review &editing. Baptiste Julien : investigation, visualization, writing - review & editing. David Fenning: supervison, funding acquisition, writing - review & editing. Jifeng Liu: supervision, funding acquisition, writing - review & editing, Geoffroy Hautier: supervision, funding acquisition, writing - review & editing, Andriy Zakutayev: supervision, funding acquisition, writing - review & editing, Sage R. Bauers: conceptualization, supervision, funding acquisition, visualization, writing - review & editing

## Conflict of Interest

The authors declare no conflict of interest.

# Supplementary Information for: Thin film synthesis of $SrZn_2P_2$ with $SrI_2$ post-annealing for enhanced crystallinity and optoelectronic quality

Sita Dugu[1], Shaham Quadir[1], Christopher P. Muzzillo[1], Zhenkun Yuan[2,3,4], Smitakshi Goswami [2,5], Xiaojing Hao[6], Jialiang Huang[6], Guillermo Esparza[7], Baptiste Julien[1], David Fenning[7], Jifeng Liu[2], Geoffroy Hautier[2,3,4], Andriy Zakutayev[1], Sage R. Bauers[1]

[1]Materials Sciences Center, National Laboratory of the Rockies, Golden, Colorado 80401, United States
[2]Thayer School of Engineering, Dartmouth College Hanover, NH 03755, USA
[3]Department of Materials Science and Nano Engineering, Rice University, Houston, TX, USA
[4] Rice Advanced Materials Institute, Rice University, Houston, TX, USA
[5]Department of Physics and Astronomy, Dartmouth College, Hanover, NH 03755, USA
[6]Australian Centre for Advanced Photovoltaics, School of Photovoltaic and Renewable Energy Engineering, University of New South Wales, Sydney, New South Wales, Australia
[7]Department of Chemical and Nano Engineering, University of California San Diego, CA 92093, US

Table of Contents:

**Additional Structural Data**

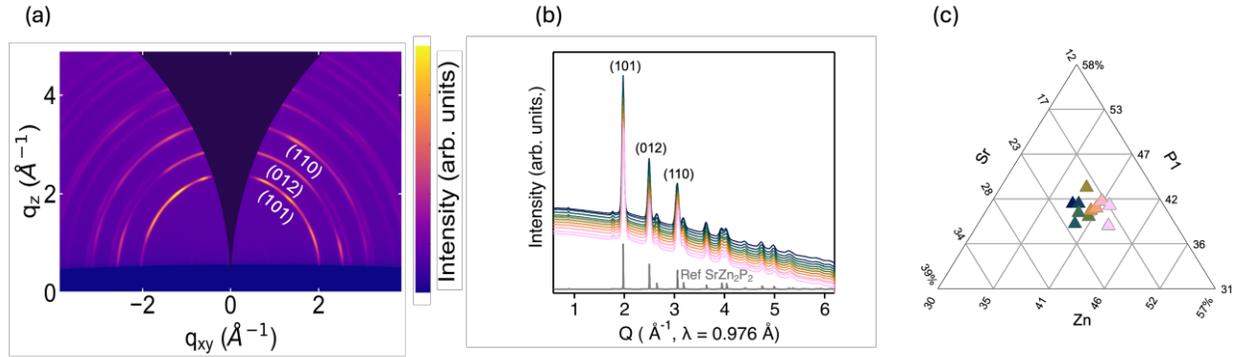

Figure S1(a) Diffraction image of GIWAXS pattern of figure 1b. (b) GIWAXS pattern of 11 different compositional point if SrZn2P2 film. (c) Elemental composition of corresponding patterns.

Table S1: Comparison of calculated and expeimental Raman modes.

| Mode | Calculated (cm$^{-1}$) | Measured (cm$^{-1}$) |
|------|------------------------|----------------------|
| Eg   | 81                     | 94                   |
| A1g  | 136                    | 135                  |
| Eg   | 276                    | 267                  |
| A1g  | 299                    | 313                  |

# X-Ray Photoelectron Spectroscopy (XPS) Analysis

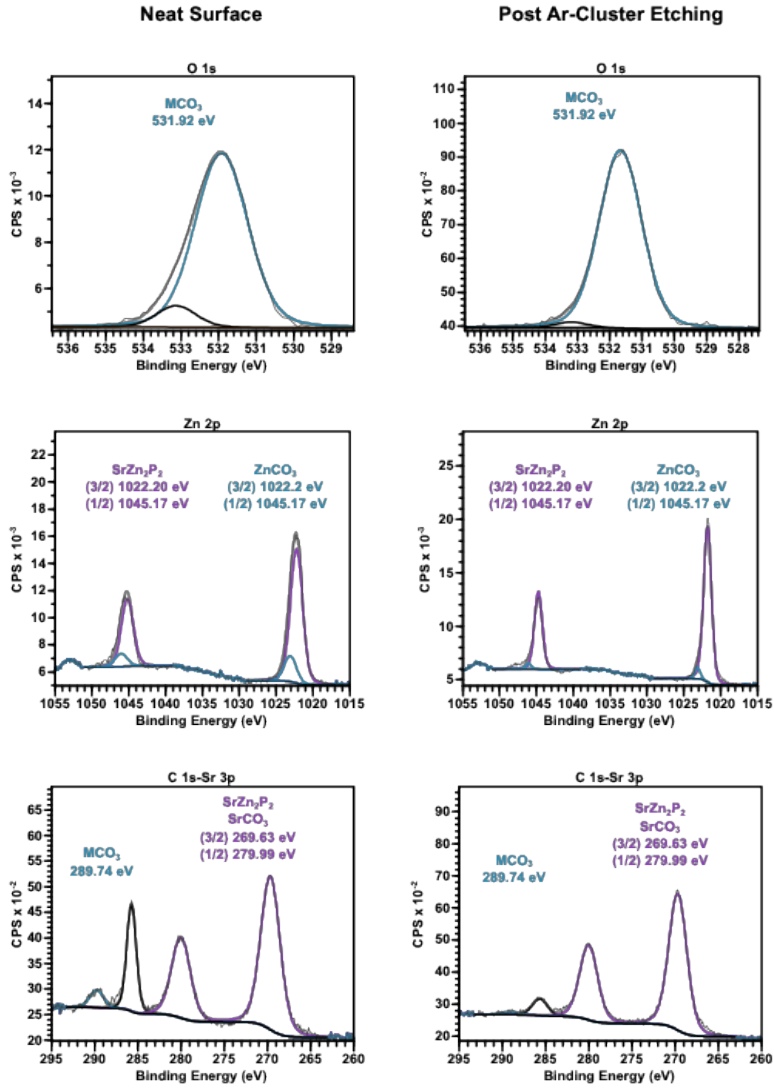

Figure S2: XPS spectra for neat surface (left panels) and post Ar-cluster etching (right panels) of $SrZn_2P_2$ films.

## Optical Analysis

The complex index of refraction is defined as $N = n + ik$. $\alpha$ is related to $k$ through $\alpha = 4\pi k/\lambda$, where $\lambda$ is the wavelength of the incident light. $\alpha$ value is found to be 10025 for wavelength 704.54 nm, close to the band energy of the system. n, k, and $\alpha$ data are obtained from $\Psi$ and $\Delta$ data. Experimental $\Psi$ and $\Delta$ data were fit using a B-spline model, yielding a mean squared error (MSE) of 5.051. The experimental $\Psi$ and $\Delta$ curve with fit is shown in Figure S2(a). The model estimates a thickness of 647 nm, and roughness 17 nm. At a wavelength 720 nm, the extracted $n$ and $k$ values are 3.1 and 0.049, respectively.

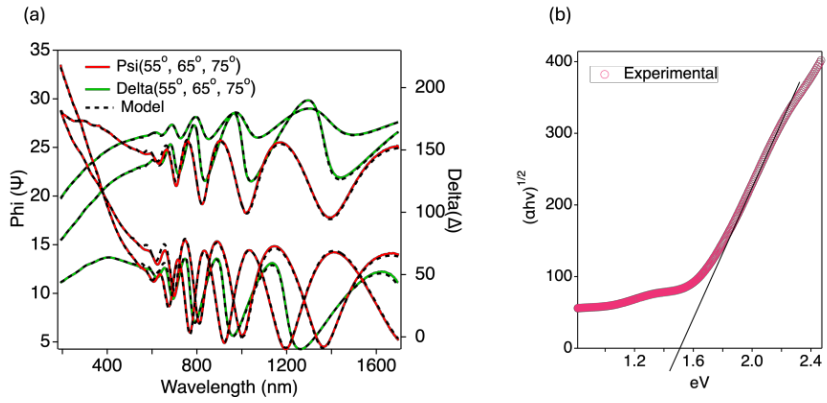

Figure S3: (a) Phi and Delta versus wavelength measured at angle 55°, 65° and 75°. (b) Tauc plot to obtain indirect band gap.

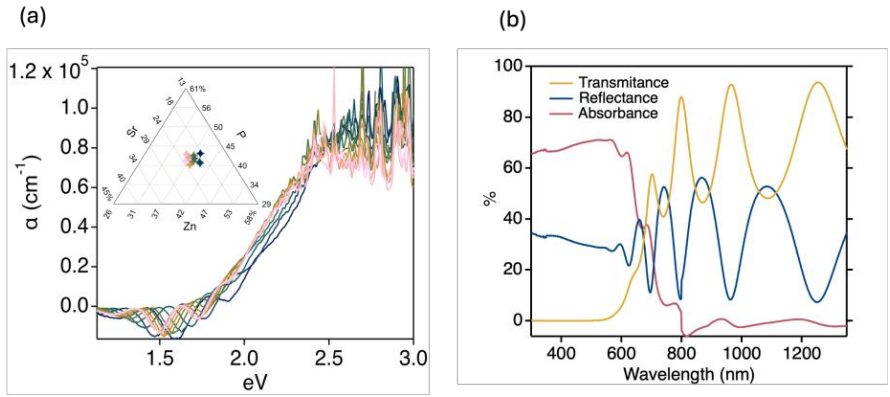

Figure S4: (a) Absorptivity of slightly off-stoichiometric $SrZn_2P_2$ (Inset shows compositions with corresponding color). (b) Transmittance, absorbance and reflectance of stoichiometric film.

Energy Dispersive X-ray Spectroscopy (EDS) data

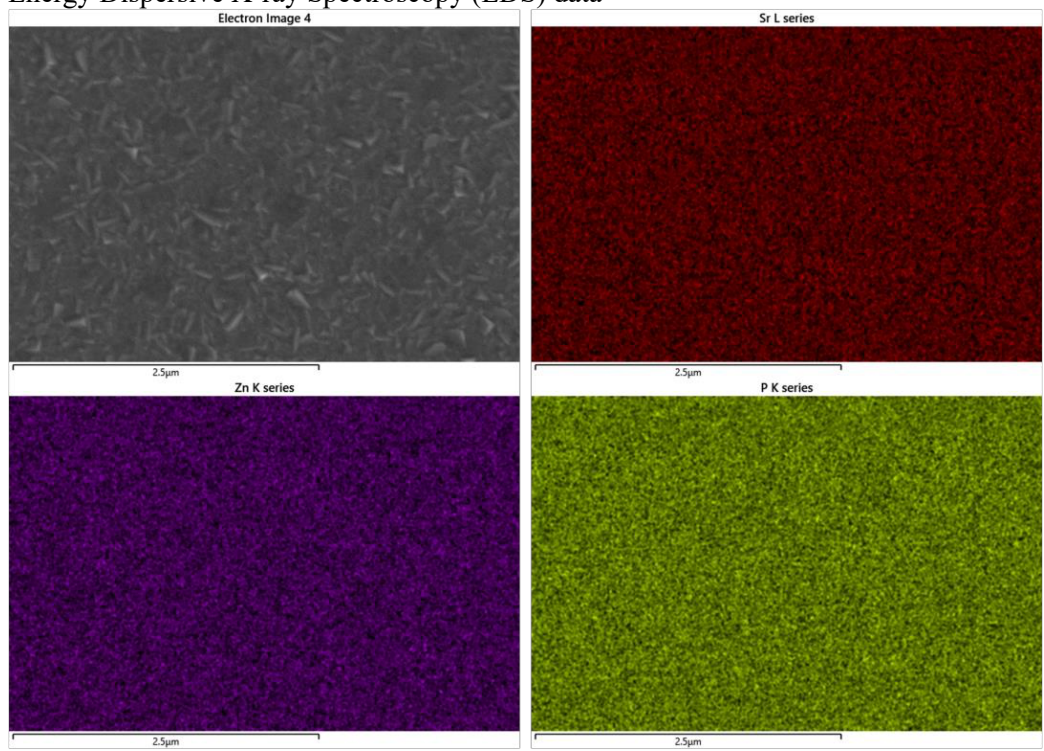

Figure S5: EDS elemental maps from the same region of the film CL measurements were collected.

**Annealing Studies**

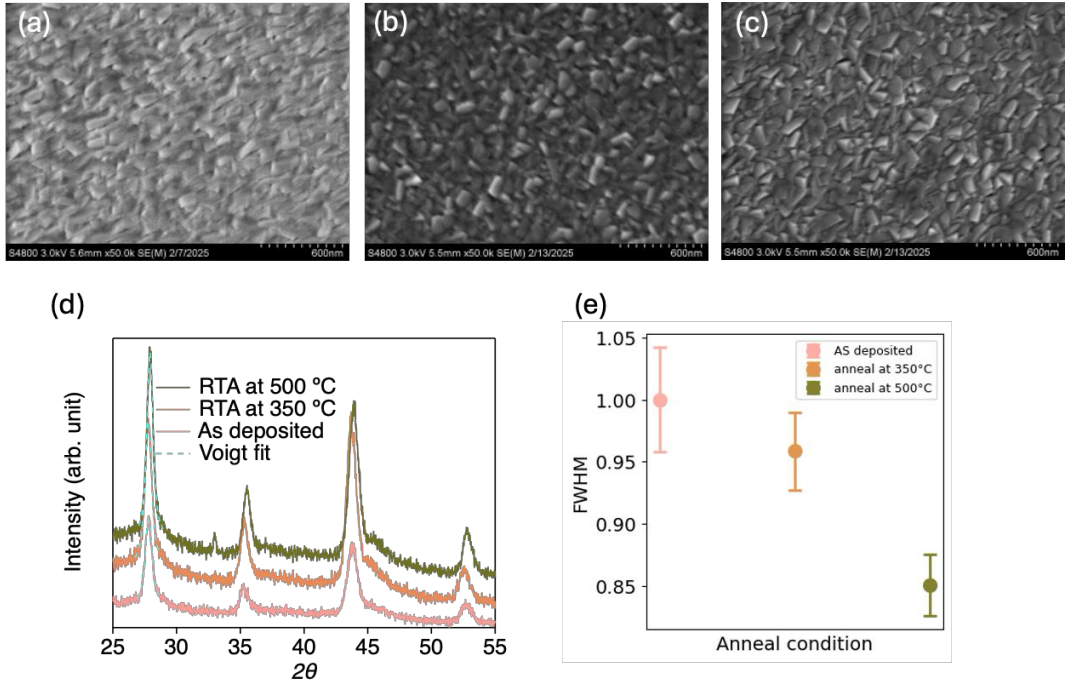

Figure S6: Plan view SEM micrographs of SrZn$_2$P$_2$ film for (a) as-deposited and rapid thermal anneal at (b) 350°C and at (c) 500°C. (d) Diffraction pattern of these three films. (e) Normalized FWHM for these three films.

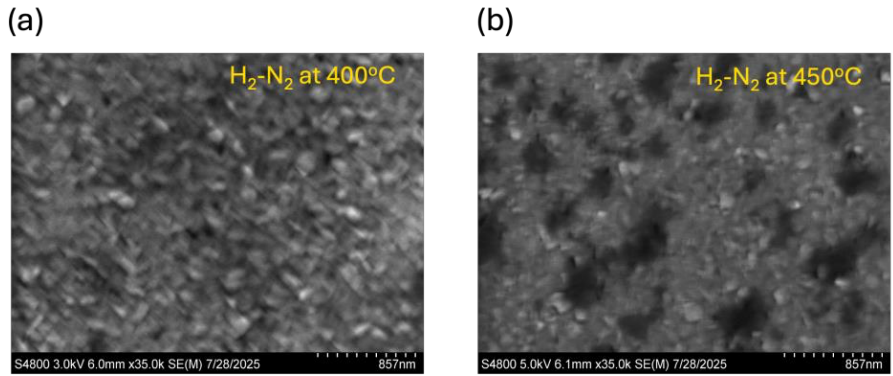

Figure S7: Plan view SEM micrographs of film annealed in H$_2$-N$_2$ environment at (a) 400°C and (b) 450°C.

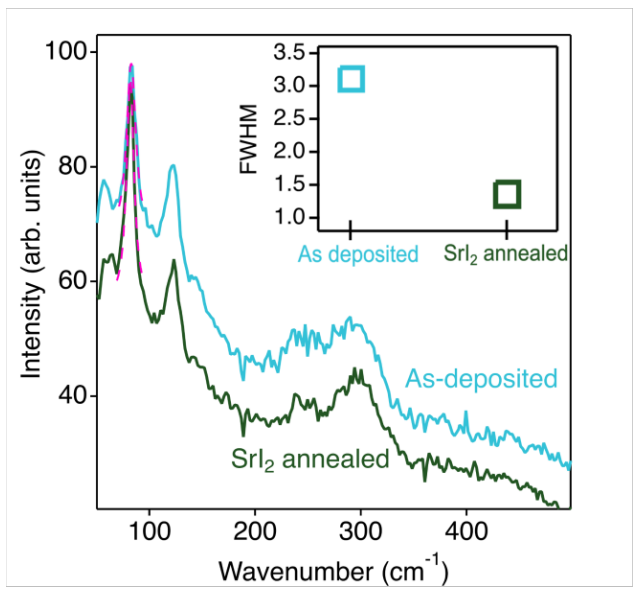

Figure S8: Raman spectra of SrZn$_2$P$_2$ film for before and after SrI$_2$ treatment at 450°C. Pink dashed curve is Voigt fit. Inset shows the FWHM from this fit.

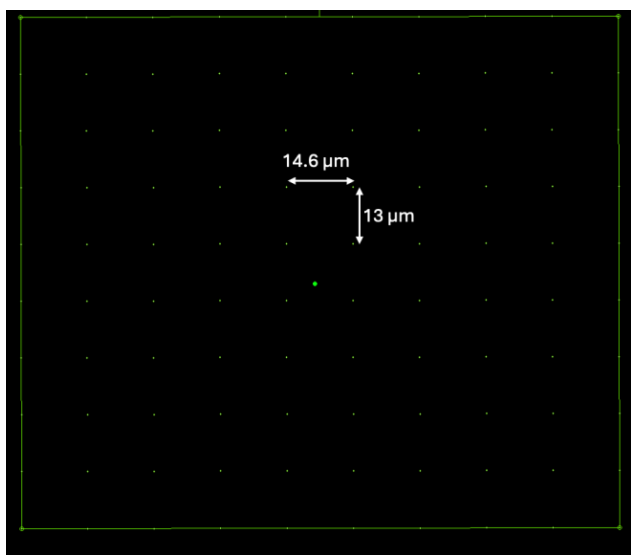

Figure S9: Schematic of the 100 points of PL scan.